\documentclass[12pt]{article}
\usepackage{a4,amssymb,verbatim,amsmath}

\newtheorem{theorem}{Theorem}[section]
\newtheorem{lemma}[theorem]{Lemma}
\newtheorem{remark}[theorem]{Remark}

\newtheorem{ex}{Example}[section]
\newenvironment{example}{\begin{ex}\rm}{ \hfill $\Diamond$ \end{ex}
        \vskip4pt}
\newtheorem{ass}{Assumption}[section]

\numberwithin{equation}{section}

\begin{document}

\newcommand{\ddt}{\partial \over \partial t}
\newcommand{\ddx}{\partial \over \partial x}
\newcommand{\ddy}{\partial \over \partial y}

\begin{center}
{\bf \Large First integrals of ordinary difference equations which
do not possess a variational formulation}
\end{center}

\begin{center}
{\bf  P. Winternitz}$^{{*}}$, {\bf V. Dorodnitsyn}$^{**}$, {\bf  E. Kaptsov}$^{\dag}$, {\bf R. Kozlov}${{\ddag}}$
\end{center}

$^{{*}}$ Centre de Recherches Math\'ematiques et D\'epartement de
math\'ematiques et de statistique, Universit\'e de Montr\'eal,
Montr\'eal, QC, H3C 3J7, Canada; {e-mail: wintern@crm.umontreal.ca}

$^{**}$
Keldysh Institute of Applied Mathematics of Russian Academy of Science,
Miusskaya Pl. 4, Moscow, 125047, Russia;
e-mail: dorod@spp.Keldysh.ru

$^{{\dag}}$  NPO Verteks, Turgenieva st. 131/1, Krasnodar, 350051, Russia;
e-mail: evgkaptsov@gmail.com

${{\ddag}}$ Department of Business and Management Science, Norwegian
School of Economics, Helleveien 30, 5045, Bergen, Norway;
e-mail: Roman.Kozlov@nhh.no

\begin{center}
29.07.2013
\end{center}

\begin{center}
{\bf  Abstract }
\end{center}

The paper presents a new method
for finding first integrals of ordinary
difference equations which do not possess Lagrangians, nor
Hamiltonians. As an example we solve a third order nonlinear
ordinary differential equation and its invariant discretization
using three first integrals obtained using this method.

\bigskip
\bigskip

\section{Introduction}  \label{introdac}

Let us first consider  a scalar $n$-th order PDE
\begin{equation}  \label{equat}
F ( x^1 , ... , x^p   ,  u ,  { \mathop{u}\limits_{1}}  ,  {
\mathop{u}\limits_{2}} ,..., { \mathop{u}\limits_{n}} )=0  ,
\end{equation}
where
$$
{ \mathop{u}\limits_{1}} := \{ u_i \} =  \left\{   {\partial u \over
\partial {x^i}  } \right\} ,
\quad ... , \quad { \mathop{u}\limits_{k}} := \{ u_{i_1 ... i_k} \}
=  \left\{ {\partial ^k u  \over \partial x^ {i_1}   ... \partial x^
{i_k} } \right\} ,
 \quad ... ,  \quad i =  1, ..., p.
$$

Let $L$ be a linear operator
\begin{equation}  \label{sym}
L = \sum _{k = 0 } ^{\infty}   F_{{u_{i_1...i_k}}}  D_{i_1} \cdots  D_{i_k}  ,
\end{equation}
where
$$
F_{{u_{i_1...i_k}}} = { \partial  F \over \partial u_{i_1...i_k} }  ,
\qquad
D_i= {\partial {}\over \partial {x^i}}
+ u _{ij}{\partial {}\over \partial {u _j}}
+ u _{ijl}{\partial {}\over \partial {u _{jl}}}
+ ...,
$$
then the adjoint operator is given by the relation
\begin{equation}  \label{adjointi}
L^*  v
=  {\delta {} \over \delta  u  }( v  F )
= \sum _{k = 0 } ^{\infty}   (-1)^k D_{i_1} \cdots  D_{i_k}
(  v  F_{u_{{i_1...i_k}}})  .
\end{equation}
It defines the adjoint equation
\begin{equation}  \label{adjointN}
F^*    =  {\delta {} \over \delta  u   } (vF) = \sum _{k = 0}
^{\infty}
 (-1)^k D_{i_1}...D_{i_k}(v F_{u _{{i_1...i_k}}})=0 .
\end{equation}

The basic operator identity  (which is probably due to Lagrange, see
for example~\cite{Dem}, Eq.~(2.75) on p.~80) is the following
\begin{equation}  \label{ident0}
v L  w  - w  L^* v  = D_i C^i,
\end{equation}
where $v$  and $w$ are some functions of ${\bf x}  = ( x^1 , ... , x^p ) $, $u$ and finite
number of derivatives of $u$. Here
\begin{equation}  \label{conslaw}
C^i
= \sum _{k = 0 } ^{\infty}   D_{i_1} \cdots  D_{i_k} ( w )
{\delta {} \over \delta  u _{i i_1...i_k} }  ( v F)  ,
\end{equation}
where
$$
{\delta {} \over \delta  u _{i i_1...i_k} } =   \sum _{s = 0 }
^{\infty} (-1)^s D_{i_1} \cdots  D_{i_s} { \partial  \over \partial
u _{{i i_1...i_k} i_1...i_s} }
$$
are higher order Euler-Lagrange operators.

Consider  Lie symmetries~\cite{Ovs1, Ibr1, Olver1}
\begin{equation}  \label{symmetr}
X =  \xi  ^i  {\partial \over \partial  x ^i  }
+   \eta  {\partial \over \partial  u  }
+  \sum _{s = 1} ^{\infty}  \zeta _{i_1 ... i_s }  {\partial \over \partial  u  _{i_1 ... i_s }   }  ,
\end{equation}
where $\xi  ^i $ and $   \eta$  are some functions of ${\bf x} $, $u$ and finite
number of derivatives of $u$ and
$$
\zeta _{i_1 ... i_s } = D _{i_1} ... D_{i_s }  (  \eta - \xi ^i u_ i
) + \xi ^i u_{i {i_1 ... i_s }} .
$$
To each Lie symmetry~(\ref{symmetr}) there corresponds the canonical
symmetry (evolutionary vector field)
\begin{equation}
\bar{X}
=   \bar{\eta}   {\partial \over \partial  u  }
+  \sum  _{s = 1} ^{\infty}  \bar{\zeta} _{i_1 ... i_s }  {\partial \over \partial  u  _{i_1 ... i_s }   }  ,
\end{equation}
 where
$$
\bar{\eta}  = \eta  - \xi ^i u _ i  , \qquad \bar{\zeta} _{i_1 ...
i_s } = D _{i_1} ... D_{i_s } (  \bar{\eta} )  .
$$

The identity~(\ref{ident0}) can be used to link symmetries of the
differential equation~(\ref{equat}), solutions of the corresponding
adjoint equation~(\ref{adjointN}) and conservation laws. Notice that
adjoint equation is always linear for $v$  (if $u$ is known).

Choosing  $w = \bar{\eta} =  \eta  - \xi ^i  u_i  $ in
(\ref{ident0}), we obtain identities
\begin{equation}
v \bar{X} F = \bar{\eta} F^* + D_i C^i
\end{equation}
and
    \begin{equation}   \label{id2}
v  X  F = v \xi  ^i D_i ( F ) +   \bar{\eta} F^* + D_i C^i  ,
\end{equation}
where
\begin{equation}  \label{conslaw2}
C^i
= \sum _{k = 0 } ^{\infty}   D_{i_1} \cdots  D_{i_k} (  \bar{\eta}  )
{\delta {} \over \delta  u _{i i_1...i_k} }  ( v F)  .
\end{equation}

One can formulate the following theorem based on the Lagrange
identity:

\begin{theorem}
The system of equations~(\ref{equat}),(\ref{adjointN}) possesses the
following conservation law
\begin{equation}  \label{cons}
D_i C^i|_{(\ref{equat}),(\ref{adjointN})} =0
\end{equation}
for each Lie symmetry~(\ref{symmetr}) of the differential equation~(\ref{equat})
and for each solution of the adjoint equation~(\ref{adjointN}).
\end{theorem}

Since we are interested in solving~(\ref{equat}) we need
conservation laws for this equation alone, without using solutions
of the adjoint equation~(\ref{adjointN}). There is a way to get rid
of the adjoint variable $v$ as suggested by
Ibragimov~\cite{ibr11a},\cite{Ibr}:

\begin{theorem}    \label{main0}
Let the adjoint equation~(\ref{adjointN}) for $v$ be satisfied for
all solutions $u$ of the differential equation~(\ref{equat}) upon a
substitution
\begin{equation}   \label{subst0}
v = \varphi ( x^1, ... x^p  ,  u ,  { \mathop{u}\limits_{1}}  ,  {
\mathop{u}\limits_{2}} , ...  ) , \qquad \varphi    {\not\equiv} 0 .
\end{equation}
Then, any Lie symmetry~(\ref{symmetr}) of the equation~(\ref{equat})
leads to the conservation law~(\ref{cons}),
where $v$ and its derivatives should be eliminated via equation~(\ref{subst0})
and its differential consequences.
\end{theorem}

The purpose of this note is to present discrete counterparts of
these results for ordinary difference equations. We do not assume
Lagrangian or Hamiltonian formulation of the equation~(\ref{equat}).
A discrete analog of the Noether theorem~\cite{Noe1}, which provides
conservation laws for difference ODEs and PDEs,  was developed
in~\cite{[66], [79], [45], [87]}. Discrete Hamiltonian equations
were considered in~\cite{[85], [75], [39]}.

\section{Scalar ODEs}  \label{sectionODE}

In this section we specify the results of Section~\ref{introdac} for
scalar ordinary differential equations (ODEs) of order~$n$
\begin{equation}  \label{ode}
F  ( x,  u  ,  \dot{u}   , \ddot{u}   , ... , u^{(n)}   ) = 0,
\end{equation}
which possesses Lie point symmetries
\begin{equation}  \label{symmetry}
X =  \xi(x,u)  {\partial \over \partial  x  }
+   \eta(x,u)  {\partial \over \partial  u  }
+ \sum _{s = 1} ^{\infty}  \zeta _{s
} {\partial \over \partial  u^{(s) }   }.
\end{equation}

In this case we have total differentiation
$$
D =  {\partial \over \partial  x  } +  \dot{u}   {\partial \over
\partial  u  } +  \ddot{u}   {\partial \over \partial  \dot{u}  } +
... +  u^{(k+1)}   {\partial \over \partial  u^{(k)}    } + ... .
$$
The variational operator is
$$
{  \delta  \over \delta   u  }
=     {\partial  \over \partial  u  }
- D    {\partial   \over \partial  \dot{u}  }
+ D ^2     {\partial   \over \partial  \ddot{u}  }
+ ...
+  (-1) ^{k}  D^k    {\partial   \over \partial  u^{(k)}    }
+ ...
$$
and higher Euler--Lagrange operators are
$$
{  \delta  \over \delta   u ^{(i)}  } =     {\partial  \over
\partial  u^{(i)}  } - D    {\partial   \over \partial   u^{(i+1)} }
+ D ^2     {\partial   \over \partial  u^{(i+2)}  } + ... +  (-1)
^{k}  D^k    {\partial   \over \partial  u^{(i+k)}    } + ...
$$

In this case identity~(\ref{id2}) takes the form
\begin{equation}     \label{identity2}
v    X  ( F )  =   v \xi D( F)  +  \bar{\eta}  F^*   + D ( I )   ,
\end{equation}
where
\begin{equation}  \label{firstint1}
I = \sum_{i = 0} ^{n-1} D^i( \bar{\eta} ) {\delta \over \delta u
^{(i+1)} } ( v F ) , \qquad \bar{\eta}  = \eta - \xi \dot{u} .
\end{equation}

Theorem~\ref{main0}  takes the following form.

\begin{theorem}   \label{main}
Let the adjoint equation
\begin{equation} \label{adjoint}
F^* = {  \delta  \over \delta   u  } ( v F )
=  v     {\partial F \over \partial  u  }
- D  \left( v   {\partial  F \over \partial  \dot{u}  } \right)
+ ...
+  (-1) ^{n}  D^n   \left( v  {\partial  F \over \partial  u^{(n)}    } \right)
= 0
\end{equation}
be satisfied
for all solutions of the original ODE~(\ref{ode})
upon a substitution
\begin{equation}  \label{subst}
v = \varphi ( x,u , \dot{u}, \ddot{u} , ... , u ^{(n-1)} ),
\qquad
\varphi    {\not\equiv} 0 .
\end{equation}
Then, any Lie point symmetry~(\ref{symmetry}) of the
equation~(\ref{ode}) leads to a first integral~(\ref{firstint1}),
where $v$ and its derivatives should be eliminated via
equation~(\ref{subst}) and its differential consequences.
\end{theorem}

First integrals $I$, given by~(\ref{firstint1}), can depend on
$u^{(n)}$ as well as higher derivatives. We will call such
expressions {\it higher} first integrals. It is reasonable to use
the ODE~(\ref{ode}) and its differential consequences to express
these first integrals as functions of the minimal set of variables,
i.e.,  in the form $ \tilde{I}  (x, u, \dot{u} , ..., u^{(n-1)} )$.

\begin{example}  \label{mainexample}

Let us consider the ODE~\cite{Pavel}
\begin{equation}  \label{third}
 F =  { 1 \over \dot{u} ^2 } \left( \dot{u}  \dddot{u} - { 3 \over 2 }  \ddot{u} ^2  \right)  =
 0,
\end{equation}
which admits symmetries
\begin{equation}   \label{part01}
X_1 = {\frac{ \partial }{\partial u}} , \qquad X_2 = u {\frac{
\partial }{\partial u}} , \qquad X_3 = u^2 {\frac{ \partial
}{\partial u}} ,
\end{equation}
\begin{equation}   \label{part02}
X_4 = {\frac{ \partial }{\partial x}} ,
\qquad X_5 = x {\frac{ \partial }{\partial x}} , \qquad X_6 = x^2
{\frac{
\partial }{\partial x}} .
\end{equation}

The adjoint equation~(\ref{adjoint}) takes the form
\begin{equation}  \label{adjoint0}
F^* = - {  \dddot{v}  \over \dot{u} }    =   0 .
\end{equation}
This linear adjoint equation has three independent solutions  of the
form $v = v (x) $:
\begin{equation}   \label{indep1}
 v_a = 1  , \qquad   v_b = x,  \qquad  v_c  = x^2  .
\end{equation}
Using these three solutions and six
symmetries~(\ref{part01}),(\ref{part02}), one can find $ 3 \times 6
= 18$ first integrals, some of which can be trivial. Among
non-trivial first integrals we chose three independent ones:
\begin{equation}   \label{inregra}
\tilde{I}_{1a} =    { \ddot{u}^2 \over 2 \dot{u}^3 }  , \qquad
\tilde{I}_{2a} =  { u  \ddot{u}^2 \over 2  \dot{u}^3 } - { \ddot{u}
\over \dot{u} }  , \qquad \tilde{I} _{1b} =  { x \ddot{u}^2 \over 2
\dot{u}^3 } + { \ddot{u} \over  \dot{u}^2 } .
\end{equation}
The notation  $\tilde{I}_{j \alpha }$ means that this integral
corresponds to symmetry $X_j$ and solution $v_{\alpha}$ of the
adjoint equation~(\ref{adjoint0}). Setting these integrals equal to
constants and eliminating $\dot{u}$ and $\ddot{u}$ from
~(\ref{inregra}),  we obtain two families of solutions (generic and
degenerate)

\begin{equation}  \label{odesolution}
u (x) = { 1 \over C_1 x + C_2 } +  C_3 \qquad \mbox{and} \qquad u(x)
= C_1 x + C_2 ,
\end{equation}
where $C_1 \neq 0$, $C_2$ and $C_3$ are constants expressed in terms
of the first integrals.

\end{example}

\section{Symmetry--preserving discretization of scalar ODEs and first integrals of
the difference schemes }   \label{discretization}

In this section we are interested in dicretizations of the scalar
ODE~(\ref{ode}). For the discretization of an ODE of order $n$ we
need a difference stencil with at least $n+1$ points. We will use
precisely $n+1$ points, namely, points $    x _m $, ...,  $ x_{m+n}
$. These points are not specified in advance and will be defined by
an additional mesh equation~\cite{[87]}.

As a discretization we will consider a discrete equation on $n+1$ points
\begin{equation} \label{difference3}
F  ( x _m ,  u_m , x_{m+1} ,  u_{m+1}  ,  ... , x_{m+n} ,  u_{m+n}   ) = 0 ,
\end{equation}
which is considered on the mesh
\begin{equation}  \label{mesh}
 \Omega   ( x _m ,  u_m , x_{m+1} ,  u_{m+1}   , ... , x_{m+n} ,  u_{m+n}   ) = 0 .
\end{equation}
These two equations form the difference system to be used. In the
continuous limit the first equation goes into the original ODE and
the second equations turns into  an identity (for example, $0=0$).

The Lie point symmetry generator is the same as in the continuous
case
\begin{equation}  \label{symmetry3}
X
= \xi  (x,u) {\partial \over \partial  x  }
+ \eta (x,u) {\partial \over \partial  u  }
\end{equation}
but its prolongation  to the points of the difference stencil is
\begin{equation}
X
= \xi  _{m} {\partial \over \partial  x _{m} }
+ \eta _{m} {\partial \over \partial  u _{m} }
+ ...
+ \xi  _{m+n} {\partial \over \partial  x _{m+n} }
+ \eta _{m+n} {\partial \over \partial  u _{m+n} } ,
\end{equation}
where $  \xi  _{k} = \xi ( x _{k} , u _{k} ) $ and $
 \eta  _{k} = \eta ( x _{k} , u _{k} ) $.

It is helpful  to introduce backwards (left) shift operator $S_-$:
$$
S_- (m)= m-1, \quad  S_- (u _m) =  u _{m-1} , \quad  S_- (x_ m) =  x
_{m-1} .
$$
Discrete variational operators are defined by the relation
$$
\delta \sum _m {\cal F}  ( m, x _m ,  u_m , x_{m+1} ,  u_{m+1}  ,  ... ,
x_{m+n} ,  u_{m+n}   )
$$
$$
= \sum _m  \left(  \delta u_m   \sum _{k = 0} ^{\infty} S_- ^k
{\partial \over \partial  u _{m+k} } + \delta x_m  \sum _{k = 0}
^{\infty} S_- ^k   {\partial \over \partial  x _{m+k} } \right) {\cal F}  (
m, x _m ,  u_m , x_{m+1} ,  u_{m+1}  ,  ... , x_{m+n} ,  u_{m+n} ) .
$$
This provides us with two operators
\begin{equation}   \label{variational3}
{ \delta   \over \delta u_m  }
= \sum _{k = 0} ^{\infty} S_- ^k   {\partial \over \partial  u _{m+k} }  ,
\qquad
{ \delta   \over \delta x_m  }
= \sum _{k = 0} ^{\infty} S_- ^k   {\partial \over \partial  x _{m+k} }  .
\end{equation}
We suppose ${\cal F} \rightarrow 0 $ sufficiently fast  when $ m
\rightarrow \pm \infty$ so that the difference functional is well
defined.  Note that these operators are given for the
scheme~(\ref{difference3}),(\ref{mesh}) with arbitrary $n$. To the
system of difference equations~(\ref{difference3}),(\ref{mesh})
there correspond the adjoint equations
\begin{equation}  \label{adjoint3}
F ^* = { \delta   \over \delta u_m  } ( v_m F + w_m \Omega ) = 0 ,
\qquad
\Omega ^* = { \delta   \over \delta x_m  } ( v_m F + w_m \Omega ) =  0  ,
\end{equation}
which are always linear for the adjoint variables $ v_m $ and $ w_m
$. Let us fix the value of index $m$, which corresponds to the left
point in the equations~(\ref{difference3}),(\ref{mesh}), and define
{\it higher}  discrete Euler--Lagrange operators
\begin{equation} \label{higher3}
{ \delta    \over \delta  u  _{m (j)}  }
= \sum _{k = 0 } ^{\infty}  S_- ^{k}  { \partial   \over \partial u  _{m+j+k}  }  ,
\qquad
{ \delta    \over \delta  x  _{m (j)}   }
= \sum _{k = 0 } ^{\infty}  S_- ^{k}  { \partial   \over \partial x  _{m+j+k}  }  .
\end{equation}

\begin{lemma}  \label{ideintity1} {\bf (Main identity)}
The following operator identity holds
\begin{equation}     \label{ideintity1a}
v _m   X ( F  )  +  w _m    X ( \Omega  )
=  \eta  _m   F^* +  \xi _m  \Omega ^*   +   ( 1 - S_- ) J  ,
\end{equation}
where
\begin{equation}
J = \sum _{j = 1 } ^{n}
\left(
\xi  _{m+j}   { \delta   \over \delta x_{m (j)}   }
+ \eta  _{m+j}   { \delta   \over \delta u_{m (j)}   }
\right)
 ( v  _m F   +  w_m  \Omega )  .
\end{equation}
\end{lemma}

The identity can be proven by direct verification. From the identity
we obtain the following result.

\begin{theorem}   \label{result3}
{\bf (Main result for discretized ODEs)} Let the adjoint
equations~(\ref{adjoint3}) be  satisfied  for all solutions of the
original equations~(\ref{difference3}),(\ref{mesh}) upon a
substitution
\begin{equation}  \label{substitition3}
\begin{array}{l}
v_m = \varphi _1 ( m,x_m,u_m , ...,  x_{m+n-1},u_{m+n-1} ) , \\
\\
w_m = \varphi _2 ( m,x_m,u_m , ...,  x_{m+n-1},u_{m+n-1} ) , \\
\end{array}
\qquad
\qquad
\varphi _1    {\not\equiv} 0
\quad
\mbox{or}
\quad
\varphi _2  {\not\equiv} 0 .
\end{equation}
Then, any Lie point symmetry~(\ref{symmetry3}) of the equations~(\ref{difference3}),(\ref{mesh})
leads to first integral
\begin{equation}    \label{f_int_ODE}
J = \sum _{j = 1 } ^{n}
\left(
\xi  _{m+j}   { \delta   \over \delta x_{m (j)}   }
+ \eta  _{m+j}   { \delta   \over \delta u_{m (j)}   }
\right)
 ( v  _m F   +  w_m  \Omega )
 ,
\end{equation}
where $v_m$, $w_m$, ...,   $v_{m-n}$, $w_{m-n}$  should be eliminated
by means of Eqs.~(\ref{substitition3}) and their shifts to the left.
\end{theorem}

First integrals $J$ which  depend on more than $n$ points can always
be expressed as  $\tilde{J} ( m, x_m, u_m , ..., x_{m+n-1},u_{m+n-1}
) $ with the help of the equations~(\ref{difference3}),(\ref{mesh}).

\begin{example}

Let us return to the ODE~(\ref{third}). As a discretization we
consider an invariant scheme which consists of invariant
discretization of the ODE
\begin{equation}   \label{discrete3e1}
F =  { u_{m+3} -  u_{m+1}  \over x_{m+3} -  x_{m+1}  } { u_{m+2} -
u_{m}  \over x_{m+2} -  x_{m}  } - { u_{m+3} -  u_{m+2}  \over
x_{m+3} -  x_{m+2}  }
{ u_{m+1} -  u_{m}  \over x_{m+1} -  x_{m}  }   = 0
\end{equation}
and invariant mesh
\begin{equation}   \label{discrete3e2}
\Omega =  {  ( x_{m+3} -  x_{m+1} ) (  x_{m+2} -  x_{m} )   \over (
x_{m+3} -  x_{m+2} ) (  x_{m+1} -  x_{m} )  }  - K     = 0   , \quad
K \neq 0,
\end{equation}
which was introduced in~\cite{Pavel}. The scheme was constructed so
as to admit all six symmetries~(\ref{part01}),(\ref{part02}).

It is convenient to rewrite the scheme as
\begin{equation}     \label{discrete2b1}
\tilde{F} = {  ( u_{m+3} -  u_{m+1} ) (  u_{m+2} -  u_{m} )   \over
             ( u_{m+3} -  u_{m+2} ) (  u_{m+1} -  u_{m} )  }  - K     = 0       ,
\end{equation}
\begin{equation}     \label{discrete2b2}
\Omega =  {  ( x_{m+3} -  x_{m+1} ) (  x_{m+2} -  x_{m} )   \over
             ( x_{m+3} -  x_{m+2} ) (  x_{m+1} -  x_{m} )  }  - K     = 0   .
\end{equation}
In this specific example the difference system  splits into two
similar independent equations, which can be considered separately.

The adjoint equations~(\ref{adjoint3}) take the form
\begin{equation}    \label{adjointm1}
\tilde{F}^*  = - v_{m}  + (K-1)  v_{m-1}   + (1-K) v_{m-2}  +  v_{m-3} = 0 ,
\end{equation}
\begin{equation}    \label{adjointm2}
\Omega ^*  = - w_{m}  + (K-1)  w_{m-1}   + (1-K) w_{m-2}  +  w_{m-3} = 0 .
\end{equation}

\begin{enumerate}

\item

It is easy to find solutions $  v_{m} = v_{m} (m) $, $  w_{m} =  0
$. We restrict ourselves  to the simplest case $K= 4$ (the other cases
will be considered elsewhere). There are three independent solutions of the adjoint equation
$$
v_m ^a = 1 ,  \qquad     v _m ^b  = m , \qquad   v _m ^c = m^2   .
$$

Applying Theorem~\ref{result3} for these solutions and symmetries
(\ref{part01}), we get $ 3 \times 3 = 9 $ first integrals. Here we
present three independent ones:
$$
\tilde{J}_{1a} = 2  \left(
{ 4 \over u_{m+2} - u_{m} }
- { 1 \over u_{m+2} - u_{m+1} }
- { 1 \over u_{m+1} - u_{m} }
\right) ,
$$
$$
\tilde{J}_{2a} =
2 \left(
  { 4 u_{m+2} \over u_{m+2} - u_{m} }
-  { u_{m+1} \over u_{m+2} - u_{m+1} }
 -    { u_{m+1} \over u_{m+1} - u_{m} }
 - 2
 \right) ,
$$
$$
\tilde{J}_{1b} = 2 m \left(
{ 4  \over u_{m+2} - u_{m} }
- { 1 \over u_{m+2} - u_{m+1} }
- { 1 \over u_{m+1} - u_{m} }
\right)
$$
$$
 +   \left(
- { 4  \over u_{m+2} - u_{m} }
+ { 3 \over u_{m+2} - u_{m+1} }
- {  1  \over u_{m+1} - u_{m} }
\right)  .
$$

\item

Since the equations~(\ref{discrete2b1}) and~(\ref{discrete2b2}) have
the same form for $u$ and for $x$,  we can consider solutions $
v_{m} = 0 $, $ w_{m} = w_{m} (m)  $ in the same manner and obtain
similar first integrals for the variable $x_m$.

\end{enumerate}

Finally, from six independent first integrals we obtain the solution
of the scheme as
\begin{equation}   \label{sol3a}
u_m =   {  1 \over C_1  m  + C_2   } +  C_3
\qquad
\mbox{or}
\qquad
u_m =  C_1  m  + C_2
\end{equation}
and for the mesh points
\begin{equation}   \label{sol3b}
x_m =   {  1 \over C_4  m  + C_5   } +  C_6
\qquad
\mbox{or}
\qquad
x_m =  C_4  m  + C_5   ,
\end{equation}
 where  $C_1 \neq 0$,   $C_2$, $C_3$,  $C_4 \neq 0$,   $C_5$ and $C_6$ are constants related to
the first integrals.

\begin{remark}
Let us note that  the solution~(\ref{sol3a}) on the
mesh~(\ref{sol3b}) can be expressed as
\begin{equation}
u_m  (x_m) = { 1 \over \alpha x_m + \beta } + \gamma
\qquad
\mbox{or}
\qquad
u_m = \alpha x_m  + \beta ,
\end{equation}
where $\alpha \neq 0 $, $\beta $  and $ \gamma$ are constants. We note
that this solution is exactly the same as
solution~(\ref{odesolution}) of the ODE~(\ref{third}), i.e., the
scheme~(\ref{discrete3e1}),(\ref{discrete3e2}) is exact.
\end{remark}

\end{example}

{\bf Acknowledgements}

The research of
P.W. was partly supported by NSERC of Canada.
The  research of V.D. and E.K. was partly supported by research grant
No. 12-01-00940-a of Russian Fund for Base Research.  The research of
R.K. was partly supported by  the Norwegian Research Council under contract Nos. 176891/V30 and 204726/V30.

\end{document}